\documentclass{amsart}

\usepackage{amsfonts}
\usepackage{graphicx}
\copyrightinfo{2018}{American Mathematical Society}

\theoremstyle{definition}

\theoremstyle{remark}

\numberwithin{equation}{section}

\def\vec#1{\textbf{#1}} 
\def\sym{\textup{Sym}(\codonset)}
\def\codonset{\mathcal{C}}
\def\aminoset{\mathcal{X}}
\def\dimerset{\mathcal{D}}
\def\baseset{\mathcal{B}}

\def\X{\mathcal{X}} 
\def\Y{\mathcal{Y}}
\def\Cset{\mathcal{C}}
\def\basefont#1{{\tt #1}} 
 
\def\R{\mathbb{R}}

\DeclareMathOperator{\sgn}{sign}  

\def\Phi{f}

\begin{document}

\title[Sequence Dependent Geometry and the Genetic Code]{Did Sequence Dependent Geometry Influence the Evolution of the Genetic Code?}

\author{Alex Kasman}
\address{Department of Mathematics\\College of Charleston\\Charleston,
SC 29424}
\email{kasmana@cofc.edu}

\author{Brenton LeMesurier}
\address{Department of Mathematics\\College of Charleston\\Charleston,
SC 29424}

\email{lemesurier@cofc.edu}


\date{\today}

\begin{abstract}
The genetic code is the function
from the set of codons to the set of amino acids by which a DNA
sequence encodes proteins.  Since the codons
also influence the shape of the DNA molecule itself, the same
sequence that encodes a protein also has a separate geometric
interpretation.  A question then arises: How well-duplexed are these
two ``codes''?  In other words, in choosing a genetic
sequence to encode a particular protein, how much freedom does one
still have to vary the geometry (or vice versa).  A recent paper by
the first author addressed this question using two different methods.
After reviewing those results, this paper addresses the same question
with a third method: the use of Monte Carlo and Gaussian sampling methods to
approximate a multi-integral representing the mutual information of a
variety of possible genetic codes.  Once again, it is found that the
genetic code used in nuclear DNA has a slightly lower than average
duplexing efficiency as compared with other hypothetical genetic
codes.  A concluding section discusses the significance of these
surprising results.
\end{abstract}

\maketitle


The first author's talk at the AMS Special Session on the Topology of
Biopolymers explored the mathematical relationship between two
different roles that a DNA sequence serves in living cells: encoding
proteins to be produced and influencing the shape of the DNA molecule
itself.  Those results were subsequently published as a journal
article \cite{Kasman}. 
After briefly summarizing the main results of that published paper,
this article takes them a step further using a more sophisticated
approach to the numerical computation of the mutual information.  By
combining Gaussian and Monte Carlo sampling methods with a new
geometric inversion formula for computing the geometries, this new
approach provides a more reliable result which strengthens and
reconfirms the previously announced conclusions.

\section{Measuring the Efficiency of Duplexed Codes}

\subsection{A Motivating Example}
Consider the following unlikely situation: You will soon need to send a text message conveying a two letter word to your friend Georgina \textit{and} you also have to send a two letter word by text message to your friend Fred.  However, because of your restrictive data plan, you must achieve this by sending a single two character message to both of them at the same time.  

You can hope to achieve this by teaching Fred one of the two functions $f_i$ and teaching Georgina the function $g$
shown in Table~\ref{tbl:fg}. Each of those functions turns one of the integers from $0$
to $5$ into a letter and can therefore be used as a simple ``code''.
For example, since Georgina knows the function $g$ you can send her the numerical message ``$24$'' and she would interpret it as ``$g(2)g(4)=\basefont{N}\basefont{O}$''.  Alternatively, she would interpret the message ``$53$'' as the exclamation ``\basefont{O}\basefont{H}''.
Similarly, using either of the two functions $f_1$ or $f_2$, Fred could recognize the signal ``$04$'' as the greeting ``\basefont{H}\basefont{I}''.

The really interesting thing is that you could send the \textit{same} two digit message to both Georgina and Fred and they would interpret it differently.  That is the defining characteristic of \textit{duplexed codes}, that the same signal has two different interpretations.  

Let us first suppose that Fred has memorized $f_1$ and Georgina knows the code $g$.  If you wanted to send Georgina a message that will be interpreted as ``\basefont{NO}'', you have four different choices of signal which would convey that message to her and each one would mean something different to Fred.  For instance, you could send ``$04$'' which Fred will interpret as ``\basefont{HI}'' or you could send ``$25$'' which has the same interpretation for Georgina but which Fred will interpret as ``\basefont{OH}''.  In this scenario, you have the freedom to send different messages to Fred while still sending the desired message to Georgina at the same time.

In contrast, things would be different if Fred had learned $f_2$ as his code instead.  Even though you would still have a choice of four signals to send Georgina that would be interpreted as ``\basefont{NO}'', you would have not be able to separately control the message that was sent to Fred because all four of the signals that mean ``\basefont{NO}'' under the code $g$ would be interpreted as ``\basefont{HI}'' using code $f_2$.  There would be no way to send Fred the message ``\basefont{OH}'' or any other message besides ``\basefont{HI}'' if Georgina's message is to be interpreted as ``\basefont{NO}''.  Even though there is nothing wrong with the code $f_2$ on its own, there is something unfortunate about its relationship to $g$ which creates an obstruction to sending the message ``\basefont{NO}'' to Georgina while simultaneously sending the message ``\basefont{OH}'' to Fred.

\begin{table}
\begin{tabular}{c|c|c|c}
$c$&$f_1(c)$&$f_2(c)$&$g(c)$\\\hline
0&\basefont{H}&\basefont{H}&\basefont{N}\\
1&\basefont{O}&\basefont{O}&\basefont{H}\\
2&\basefont{O}&\basefont{H}&\basefont{N}\\
3&\basefont{I}&\basefont{O}&\basefont{H}\\
4&\basefont{I}&\basefont{I}&\basefont{O}\\
5&\basefont{H}&\basefont{I}&\basefont{O}\\
\end{tabular}
\caption{The functions $f_1$, $f_2$ and $g$ used in this introduction to illustrate duplexing and
  mutual information.}\label{tbl:fg}
\end{table}

Loosely speaking, we say that two codes are \textit{well-duplexed} if
such obstructions to encoding two messages simultaneously  are rare.
Conversely they are \textit{poorly-duplexed} if the choice of a
message for one recipient severely restricts the messages that can be
sent to the other recipient with the same signal.
A more rigorous and quantifiable method of determining whether two
codes are well-duplexed or poorly duplexed is by using the concept of
\textit{mutual information} that is part of the branch of mathematics
knowns as information theory.

\subsection{Duplexed Codes and Mutual Information}

Let us say that $f$ and $g$ are \textit{duplexed codes} whenever  $f:\Cset\to\X$ and $g:\Cset\to\Y$ are two functions with the same domain.
The terminology makes sense when one imagines sending a single ``signal'' $c\in\Cset$ to two recipients each of whom knows one of those two codes.  The goal of this section is to introduce a number associated to any duplexed codes which measures how much freedom you have to send different messages to one recipient even after the message for the other recipient is fixed.

For a randomly selected element $c\in\Cset$, let $P_f(x)$ denote the
probability that $f(c)=x\in\X$, $P_g(y)$ be the probability that
$g(c)=y$, and $P_{f,g}(x\hbox{ and }y)$  be the probability that both
$f(c)=x$ and $g(c)=y$.  

For example,  using
the functions defined in Table~\ref{tbl:fg} with domain $\Cset=\{0,1,2,3,4,5\}$, we see $f_1(c)=\basefont{N}$
is true for two of the six possible values of $c$ and so $P_{f_1}(\basefont{N})=2/6=1/3$.  Moreover,
$P_{f_1,g}(\basefont{H},\basefont{N})=1/6$ since the only way that
$f_1(c)=\basefont{H}$  and $g(c)=\basefont{N}$ could both be true is if $c=0$.  However, $P_{f_2,g}(\basefont{H},\basefont{N})=1/3$ since both $c=0$ and $c=2$ satisfy $f_1(c)=\basefont{H}$ and $g(c)=\basefont{N}$.

The mutual
information (measured in bits) of  the duplexed codes $f:\Cset\to \X$ and $g:\Cset\to \Y$ is defined to be\footnote{When $P_{f,g}(x\hbox{ and }y)=0$ it is understood that $\displaystyle P_{f,g}(x\hbox{ and }y)\log_2\left(\frac{P_{f,g}(x\hbox{ and }y)}{P_f(x)P_g(y)}\right)=0$.}
\begin{equation}
 M(f,g)=\sum_{y\in \Y}\sum_{x\in
  \X}P_{f,g}(x\hbox{ and }y)\log_2\left(\frac{P_{f,g}(x\hbox{ and }y)}{P_f(x)P_g(y)}\right).
\label{eqn:MI}
\end{equation}
It is easy to see that $0\leq M(f,g)$ is true for any two codes $f$ and $g$.
The minimum possible value of $0$ occurs when
$P_{f,g}(x\hbox{ and }y)=P_f(x)P_g(y)$ for all choices of $x$ and
$y$.  A familiar fact from probability theory is that the joint
probability is equal to the product of the two probabilities precisely
when the events are independent.  Indeed, the same idea applies here,
although we now interpret it in terms of the independence of the two
codes.  If the mutual information of two codes is zero then this tells us that the codes are
very well-duplexed in that the selection of a message to one
recipient does not restrict the message that can be sent to the
other. 

Since a mutual information of $0$ represents the best possible duplexing of codes, larger mutual information means that the codes are \textit{not} as well-duplexed.  
For example, we can compute that
$$
M(f_1,g)\approx0.584963\qquad\hbox{and}\qquad M(f_2,g)\approx 1.58496.
$$
for the codes $f_1$, $f_2$ and $g$ from Table~\ref{tbl:fg} in the previous section.  The combination of functions $f_2$ and $g$ is a bad choice for duplexing since if we were using those as codes for message to send Fred and Georgina then we could not separately choose a message for each recipient.  In contrast, $f_1$ and $g$ work better as a combination because even after we have chosen the message for one of the intended recipients we still have a choice of message that can be sent to the other.  This is reflected here in the fact that $M(f_1,g)<M(f_2,g)$; the mutual information when using $f_1$ is closer to zero and therefore closer to being optimal for duplexing.

\subsection{Comparisons with  Expected Values}

Let  $F:S\to\R$ be a real-valued function on the finite set $S=\{\sigma_1,\ldots,\sigma_n\}$.  Then define the expected value $E_S(F(\sigma))$ by the familiar formula
$$
E_S(F(\sigma))=\frac{1}{n}\sum_{\sigma\in S} F(\sigma).
$$
You will probably notice that this is nothing other than the \textit{mean} of the values that $F$ takes.  The terminology ``expected value'' taken from probability theory is a notion analogous to the average in the context of random variables.  The way to interpret it here is to imagine an experiment in which you randomly select an element $\sigma$ from $S$ and make a measurement of it to find the value $F(\sigma)$.  Then $E_S(F(\sigma))$ is the expected value in the sense that it would be the average of the measurements after a large number of experiments.  In particular, if for a particular $\hat \sigma\in S$ one has 
$$
F(\hat \sigma)<E_S(F(\sigma))
$$
then one can say that the value of $F(\hat \sigma)$ is lower than the value one would \textit{expect} for a randomly selected element of $S$.

For example, using the functions $f_1$, $f_2$ and $g$ from the motivating example above, we can consider  the mutual information $M(f_{\sigma},g)$ as a real-valued function on the index set $S=\{1,2\}$.  Then 
$$
0.584963\approx M(f_1,g)<E_S(M(f_{\sigma},g))\approx\frac{0.584963+1.58496}{2}\approx1.0849615,
$$
 tells us that the duplexing of the code $f_1$ with $g$ is better than average for codes selected from $\{f_1,f_2\}$.  Although we already knew that in this case simply by comparing the individual mutual information values, this notation will prove useful below where we will be doing something similar but with a very large index set.

\section{A Natural Example of Duplexed Codes Associated to DNA}

\subsection{The Genetic Code}

Let $\baseset=\{\basefont{A},\basefont{C},\basefont{G},\basefont{T}\}$ be the set of DNA
bases.  
Because DNA sequences of length 2 and 3 will play special roles in
this paper, let us introduce the following terminology and notation:
The set of  \textit{dimers} (length two sequences)
is
$
\dimerset=\{b_1b_2\ :\ b_i\in\baseset\}.
$
and the set of \textit{codons} (length three sequences) is
$
\codonset=\{b_1b_2b_3\ :\ b_i\in\baseset\}
$.

A \relax{genetic code} is simply a function
$\Phi_I$ from the set $\codonset$ of codons to the set
$\aminoset$ of
amino acids (and the word ``stop''):
$$
\aminoset=\{\hbox{I,L,V,F,M,C,A,G,P,T,S,Y,W,Q,N,H,E,D,K,R,Stop}\}.
$$ The genetic code used by the nuclear DNA in humans is shown in
Table~\ref{tbl:geneticcode}, and this is the same genetic code used by
nearly all known living organisms \cite{UnivGenCode2,UnivGenCode1}.
We will refer to the particular genetic code given in
Table~\ref{tbl:geneticcode} as ``the natural genetic
code'' so as to distinguish it from other hypothetical codes that are
not found in biology but will be used for comparison later in the
paper.

\begin{table}
\centering\scriptsize
\begin{tabular}{c|l}
Codon ($c$) & Amino Acid ($\Phi_I(c)$)\\ \hline \basefont{ATT} , \basefont{ATC} , \basefont{ATA} & I
  \\ \hline \basefont{CTT} , \basefont{CTC} , \basefont{CTA} , \basefont{CTG} ,
  \basefont{TTA} , \basefont{TTG} & L \\ \hline \basefont{GTT} , \basefont{GTC} ,
  \basefont{GTA} , \basefont{GTG} & V \\ \hline \basefont{TTT} , \basefont{TTC} &
  F \\ \hline \basefont{ATG} & M \\ \hline \basefont{TGT} , \basefont{TGC} & C
  \\ \hline \basefont{GCT} , \basefont{GCC} , \basefont{GCA} , \basefont{GCG} & A
  \\ \hline \basefont{GGT} , \basefont{GGC} , \basefont{GGA} , \basefont{GGG} & G
  \\ \hline \basefont{CCT} , \basefont{CCC} , \basefont{CCA} , \basefont{CCG} & P
  \\ \hline \basefont{ACT} , \basefont{ACC} , \basefont{ACA} , \basefont{ACG} & T
  \\ \hline \basefont{TCT} , \basefont{TCC} , \basefont{TCA} , \basefont{TCG} ,
  \basefont{AGT} , \basefont{AGC} & S \\ \hline \basefont{TAT} , \basefont{TAC} &
  Y \\ \hline \basefont{TGG} & W \\ \hline \basefont{CAA} , \basefont{CAG} & Q
  \\ \hline \basefont{AAT} , \basefont{AAC} & N \\ \hline \basefont{CAT} ,
  \basefont{CAC} & H \\ \hline \basefont{GAA} , \basefont{GAG} & E \\ \hline
  \basefont{GAT} , \basefont{GAC} & D \\ \hline \basefont{AAA} , \basefont{AAG} &
  K \\ \hline \basefont{CGT} , \basefont{CGC} , \basefont{CGA} , \basefont{CGG} ,
  \basefont{AGA} , \basefont{AGG} & R \\ \hline \basefont{TAA} , \basefont{TAG} ,
  \basefont{TGA} & Stop \\ \hline
\end{tabular}
\caption{ This table defines the genetic code $\Phi_I:\codonset\to\aminoset$.
  \textit{Each} of the codons from $\codonset$ in the first column is a
  pre-image of the corresponding amino acid (or ``Stop'') in the
  second.}\label{tbl:geneticcode}
\end{table}

However, it is important to realize that there are other genetic codes
that are used by biological organisms (notably, mitochondria use a
different genetic code) and that scientists have also introduced
artificial genetic codes which nevertheless seem to function well
enough to support life
\cite{malleable1,malleable3,malleable5,malleable6,malleable7,malleable4,malleable2}.
So, there is no \textit{physical law} requiring
this to be the genetic code.  In theory, the genetic code could have
been different and it is reasonable to ask the question ``Why do
nearly all living organisms use this particular genetic code?''

There is evidence to support the hypothesis that the natural genetic
code is the result of a combination of coincidences and evolutionary
pressures (see \cite{ZY} and references therein).  For example, two
codons for the same amino acid differ only in the third base much more
frequently than would be predicted by chance if the genetic code was
to be constructed entirely randomly.  This has evolutionary advantages
in that it decreases the likelihood that a mutation or mis-pairing of
mRNA and tRNA will produce a
different protein \cite{biochem,wobble}.  It is therefore presumed
that this feature is not a coincidence but an example of the effect of
natural selection on the formation of the genetic code.

\subsection{Sequence Dependent DNA Geometry}

When shown in illustrations, DNA often looks like a perfectly straight
double-helix, a twisted ladder with ``rungs'' that are the base pairs
carrying the genetic sequence.
However, real DNA is not straight; it is bent
and twisted into compact shapes that fit into living cells.

It is perhaps not surprising that the way that a DNA molecule bends is
affected by the sequence of bases which make it up.
After all,
$\basefont{A}$, $\basefont{C}$, $\basefont{G}$, and $\basefont{T}$
in $\baseset$ are not just abstract mathematical symbols.  They
represent actual chemical structures that form the base pairs in a DNA
molecule.  Hence, the electrical repulsion and attraction between
successive ``rungs'' in the DNA ladder will vary with that sequence.

Olson et all \cite{Olson} experimentally determined the geometry of
each of the $16$ dimers $d\in\dimerset$ by repeatedly measuring the
configurations of DNA strands that were two base pairs long.  They
computed the average and standard deviation of each of the six
Hassan-Calladine dimer step parameters (see \cite{HC,Kasman}).  Their results are
shown in Table~\ref{tbl:olsondata}

\begin{table}
\scriptsize
\begin{tabular}{r|c|c|c|c|c|c}
$d$ &$\bar \Delta_1(d)\ (\hat \Delta_1(d))$ & $\bar \Delta_2(d)\ (\hat
  \Delta_2(d))$ & $\bar \Delta_3(d)\ (\hat \Delta_3(d))$ & $\bar
  \theta_1(d)\ (\hat \theta_1(d))$ & $\bar \theta_2(d)\ (\hat
  \theta_2(d))$ & $\bar \theta_3(d)\ (\hat \theta_3(d))$ \\ \hline
  \basefont{AA} & $-0.03\ (0.57)$ & $-0.08\ (0.45)$ & $3.27\ (0.22)$ &
  \ $-1.4\ (3.3)$ & $0.07\ (5.4)$ & $35.1\ (3.9)$\\ \hline \basefont{AC}
  & $0.13\ (0.59)$ & $-0.58\ (0.41)$ & $3.36\ (0.23)$ &
  $-0.1\ \ (3.1)$ & $0.7\ (3.9)$ & $31.5\ (4.2)$\\ \hline \basefont{AG} &
  $0.09\ (0.69)$ & $-0.25\ (0.41)$ & $3.34\ (0.23)$ & $-1.7\ \ (3.3)$
  & $4.5\ (3.4)$ & $31.9\ (4.5)$\\ \hline \basefont{AT} & $0\ (0.57)$ &
  $-0.59\ (0.31)$ & $3.31\ (0.21)$ & \ $0\ (2.5)$ & $1.1\ (4.9)$ &
  $29.3\ (4.5)$\\ \hline \basefont{CA} & $0.09\ (0.55)$ & $0.53\ (0.89)$
  & $3.33\ (0.26)$ & \ $0.5\ (3.7)$ & $4.7\ (5.1)$ &
  $37.3\ (6.5)$\\ \hline \basefont{CC} & $-0.05\ (0.76)$ &
  $-0.22\ (0.64)$ & $3.42\ (0.24)$ & $0.1\ \ (3.7)$ & $3.6\ (4.5)$ &
  $32.9\ (5.2)$\\ \hline \basefont{CG} & $0\ (0.87)$ & $0.41\ (0.56)$ &
  $3.39\ (0.27)$ & \ $0\ (4.2)$ & $5.4\ (5.2)$ &
  $36.1\ (5.5)$\\ \hline \basefont{CT} & $0.28\ (0.46)$ & $0.09\ (0.7)$ &
  $3.37\ (0.26)$ & \ $1.5\ (3.8)$ & $1.9\ (5.3)$ &
  $36.3\ (4.4)$\\ \hline \basefont{GA} & $-0.28\ (0.46)$ & $0.09\ (0.7)$
  & $3.37\ (0.26)$ & \ $-1.5\ (3.8)$ & $1.9\ (5.3)$ &
  $36.3\ (4.4)$\\ \hline \basefont{GC} & $0\ (0.61)$ & $-0.38\ (0.56)$ &
  $3.4\ (0.24)$ & \ $0\ (3.9)$ & $0.3\ (4.6)$ & $33.6\ (4.7)$\\ \hline
  \basefont{GG} & $0.05\ (0.76)$ & $-0.22\ (0.64)$ & $3.42\ (0.24)$ &
  $-0.1\ \ (3.7)$ & $3.6\ (4.5)$ & $32.9\ (5.2)$\\ \hline \basefont{GT} &
  $-0.09\ (0.55)$ & $0.53\ (0.89)$ & $3.33\ (0.26)$ & $-0.5\ \ (3.7)$
  & $4.7\ (5.1)$ & $37.3\ (6.5)$\\ \hline \basefont{TA} & $0\ (0.52)$ &
  $0.05\ (0.71)$ & $3.42\ (0.24)$ & \ $0\ (2.7)$ & $3.3\ (6.6)$ &
  $37.8\ (5.5)$\\ \hline \basefont{TC} & $-0.09\ (0.69)$ &
  $-0.25\ (0.41)$ & $3.34\ (0.23)$ & $1.7\ \ (3.3)$ & $4.5\ (3.4)$ &
  $31.9\ (4.5)$\\ \hline \basefont{TG} & $-0.13\ (0.59)$ &
  $-0.58\ (0.41)$ & $3.36\ (0.23)$ & $0.1\ \ (3.1)$ & $0.7\ (3.9)$ &
  $31.5\ (4.2)$\\ \hline \basefont{TT} & $0.03\ (0.57)$ & $-0.08\ (0.45)$
  & $3.27\ (0.22)$ & \ $1.4\ (3.3)$ & $0.07\ (5.4)$ &
  $35.1\ (3.9)$\\ \hline
\end{tabular}
\caption{This table shows the mean ($\bar\Delta_i$) and
  standard deviation ($\hat\Delta_i$) of each of the
  Hassan-Caladine step parameters for each dimer 
  as determined experimentally in \cite{Olson}. 
} \label{tbl:olsondata}
\end{table}

Assuming that the geometric configuration of each dimer in a longer
sequence has the same expected values and standard deviations as the
isolated dimers in those experiments, it is possible to make a similar
table for the geometric configurations associated to each of the $64$
codons in $\codonset$.  The function
\begin{equation}
\bar{g}:\codonset\to\mathbb{R}^3\label{eqn:barGamma}
\end{equation}
shown in Table~\ref{tbl:Gammabar} which associates to each codon
$c\in\codonset$ a 3-tuple of numbers $\bar{g}(c)$ which gives the
location of the center of the top of the codon in Angstroms if its
base is located at the origin and if each of the dimers takes exactly
the expected geometry according to Olson et al.  (Note: In
\cite{Kasman}, this role is played by a function
$\bar\Gamma:\codonset\to\R^6$ whose image has six components because
it has angular information as well, but for simplicity in this note we
are considering only the first three components which encode the
location of the center of the third rung and not the way it is
tilted.)

\begin{table}
\centering
\hbox{\small\begin{tabular}{c|c}
Codon ($c$) & $\bar {g}(c)$\\\hline
 \basefont{AAA} &
$(-0.0200699,-0.0157676,6.54172)$\\
 \basefont{AAC} &
$(0.507057,-0.214617,6.63862)$\\
 \basefont{AAG} &
$(0.259583,0.0783038,6.60773)$\\
 \basefont{AAT} &
$(0.435648,-0.32513,6.59098)$\\
 \basefont{ACA} &
$(0.088445,0.0062801,6.68549)$\\
 \basefont{ACC} &
$(0.544148,-0.604157,6.77731)$\\
 \basefont{ACG} &
$(0.131333,-0.110103,6.74604)$\\
 \basefont{ACT} &
$(0.521928,-0.212063,6.72169)$\\
 \basefont{AGA} &
$(0.248608,-0.0483119,6.71895)$\\
 \basefont{AGC} &
$(0.78757,-0.206103,6.71786)$\\
 \basefont{AGG} &
$(0.76381,0.0120173,6.72551)$\\
 \basefont{AGT} &
$(0.101016,0.416757,6.66124)$\\
 \basefont{ATA} &
$(0.271807,-0.439449,6.72675)$\\
 \basefont{ATC} &
$(0.483576,-0.728264,6.64251)$\\
 \basefont{ATG} &
$(0.577547,-1.03579,6.6656)$\\
 \basefont{ATT} &
$(0.350045,-0.605603,6.57626)$\\
 \basefont{CAA} &
$(0.329419,0.573579,6.58491)$\\
 \basefont{CAC} &
$(0.869997,0.394174,6.63371)$\\
 \basefont{CAG} &
$(0.610102,0.677742,6.62709)$\\
 \basefont{CAT} &
$(0.798832,0.281315,6.59152)$\\
 \basefont{CCA} &
$(0.0575601,0.353981,6.74685)$\\
 \basefont{CCC} &
$(0.532134,-0.244268,6.82172)$\\
 \basefont{CCG} &
$(0.106247,0.239328,6.8063)$\\
 \basefont{CCT} &
$(0.497544,0.14655,6.7635)$\\
 \basefont{CGA} &
$(0.0878838,0.434499,6.76845)$\\
 \basefont{CGC} &
$(0.636519,0.316015,6.75001)$\\
 \basefont{CGG} &
$(0.597463,0.531588,6.76423)$\\
 \basefont{CGT} &
$(-0.0939686,0.889059,6.72712)$\\
 \basefont{CTA} &
$(0.455558,0.20792,6.78231)$\\
 \basefont{CTC} &
$(0.698918,-0.0510585,6.6871)$\\
 \basefont{CTG} &
$(0.829989,-0.345276,6.70098)$\\
 \basefont{CTT} &
$(0.550076,0.055828,6.62638)$\\
\end{tabular}
\qquad
\begin{tabular}{c|c}
Codon ($c$) & $\bar {g}(c)$\\\hline
 \basefont{GAA} &
$(-0.162531,0.126559,6.6418)$\\
 \basefont{GAC} &
$(0.371275,-0.0603654,6.72489)$\\
 \basefont{GAG} &
$(0.11697,0.227073,6.69827)$\\
 \basefont{GAT} &
$(0.300772,-0.172859,6.68065)$\\
 \basefont{GCA} &
$(-0.133882,0.138511,6.72633)$\\
 \basefont{GCC} &
$(0.343321,-0.455178,6.81874)$\\
 \basefont{GCG} &
$(-0.0871259,0.0235416,6.78672)$\\
 \basefont{GCT} &
$(0.307084,-0.0639479,6.76442)$\\
 \basefont{GGA} &
$(0.169153,-0.187748,6.79938)$\\
 \basefont{GGC} &
$(0.710666,-0.336536,6.7972)$\\
 \basefont{GGG} &
$(0.683221,-0.11919,6.81159)$\\
 \basefont{GGT} &
$(0.0142188,0.276429,6.75535)$\\
 \basefont{GTA} &
$(0.138917,0.758201,6.73073)$\\
 \basefont{GTC} &
$(0.383801,0.499296,6.63932)$\\
 \basefont{GTG} &
$(0.520989,0.208472,6.66249)$\\
 \basefont{GTT} &
$(0.230872,0.600458,6.57895)$\\
 \basefont{TAA} &
$(0.272116,0.0943844,6.68384)$\\
 \basefont{TAC} &
$(0.812838,-0.080167,6.74664)$\\
 \basefont{TAG} &
$(0.55067,0.201059,6.73324)$\\
 \basefont{TAT} &
$(0.743802,-0.193659,6.70265)$\\
 \basefont{TCA} &
$(0.146711,0.202584,6.67568)$\\
 \basefont{TCC} &
$(0.611685,-0.405631,6.72668)$\\
 \basefont{TCG} &
$(0.194597,0.0857918,6.73151)$\\
 \basefont{TCT} &
$(0.583038,-0.0130668,6.67772)$\\
 \basefont{TGA} &
$(-0.153175,-0.669182,6.73584)$\\
 \basefont{TGC} &
$(0.384081,-0.831342,6.75681)$\\
 \basefont{TGG} &
$(0.361217,-0.613645,6.77368)$\\
 \basefont{TGT} &
$(-0.294599,-0.200635,6.69264)$\\
 \basefont{TTA} &
$(0.113284,-0.0740582,6.68867)$\\
 \basefont{TTC} &
$(0.353916,-0.33747,6.59885)$\\
 \basefont{TTG} &
$(0.478231,-0.634527,6.61433)$\\
 \basefont{TTT} &
$(0.209316,-0.226962,6.53448)$\\
 \end{tabular}}
 \medskip
 
 \caption{The expected location of the center for the third rung of
   each codon relative to the position of the first rung in Angstroms.  (See Figure~\ref{fig:codonspread}
for a visual representation of this data and 
  \cite{Kasman} for mathematical details.) }\label{tbl:Gammabar}

\end{table}

\unitlength1.25in
\begin{figure}
\begin{picture}(2.66,4.32)
\put(0,0){\makebox(0,0)[bl]{\includegraphics[width=2.66\unitlength,height=4.32\unitlength]{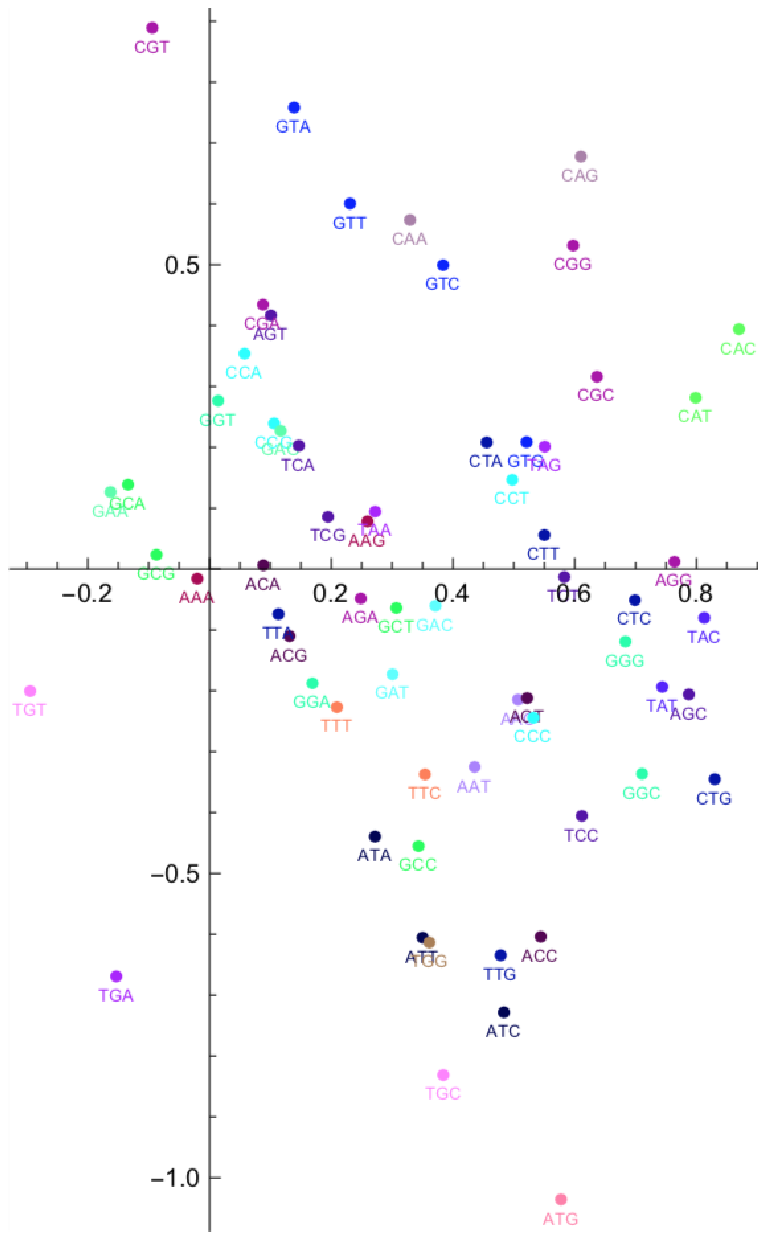}}}
\end{picture}
\caption{This graphic shows the $x$ and $y$-coordinates of the images of the function $\bar{g}(c)$ (Table~\ref{tbl:Gammabar}) for each codon $c\in\codonset$.   They show the ways that the DNA molecule carrying that sequence would likely bend.  If one was looking down at all 64 codons, each with its
  bottom rung fixed at the origin and with each dimer exhibiting its expected
  geometry (see Table~\ref{tbl:olsondata}) as it comes up out of
  the page towards you, the projections of the 
  centers of the top rungs would be located at the locations indicated
  (with axes measured in Angstroms).    }\label{fig:codonspread}
\end{figure}

Figure~\ref{fig:codonspread}  shows just the projection of $\bar{g}(c)$ onto its first two coordinates for each of the 64 codons $c\in\codonset$.  
You can imagine that a
codon (a DNA sequence of length 3) is coming straight out of the $xy$-plane
at you.  Each point in this figure represents a codon and they all
start out at the origin, but because the expected dimer
step parameters depend on the particular bases involved, by the time
they get up to their third rung they are in slightly different
positions.  In particular, the points indicate the locations of the
center of the third rung (with units given in Angstroms) if each of
the dimer step parameters takes its expected values in agreement with
the experiments of Olson et al.

As you can
see, the different codons do have slightly different expected
geometries.  It is important to realize that these small differences
can combine in dramatic ways when considering longer sequences made up
of many successive codons.  For instance,
Figure~\ref{fig:bendingdemo} shows the expected geometry for two
different DNA sequences.  Clearly, the sequence 
$$
S_1=\basefont{AAAAACGGGCAAAAACGGGCAAAAACGGGCAAAAACGGGCAAAAACGGGCAAAAACGGGC}
$$
 bends significantly more than sequence
$$
S_2=\basefont{AAGAATGGGCAGAAGCGTGCGAAGACTGGAAAGAATGGCCAGAAGCGTGCAAAAACGGGT}.
$$
So, geometrically they are quite different.  But, consider how each of
these two sequences is translated into a protein according to the
natural genetic code.  The first codon in $S_1$ (\basefont{AAA}) and
the first codon in $S_2$ (\basefont{AAG}) both encode the amino acid
\basefont{K}.  Similarly, the second codon in each encode the amino
acid \basefont{N}.  

In fact, the corresponding codons in each sequence always are mapped
by the natural genetic code to the same amino acid.  So, $S_1$ and
$S_2$ encode exactly the same protein according to the natural genetic
code, according to the function $\bar{g}$, one of them exhibits
a much greater curvature than the other.

\unitlength.7in
\begin{figure}
\begin{picture}(2.74,4.32)
\put(0,0){\makebox(0,0)[bl]{\includegraphics[width=2.74\unitlength,height=4.32\unitlength]{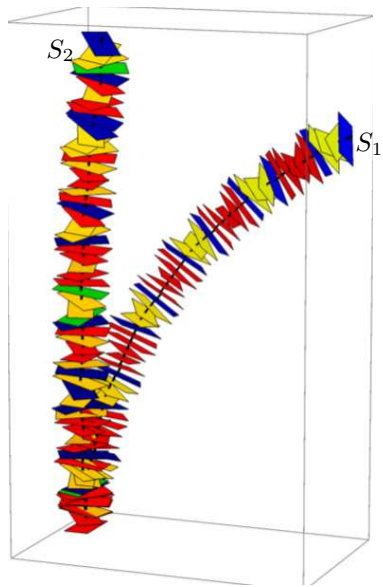}}}
\put(2.6,3.3){\makebox(0,0)[l]{$S_1$}}
\put(.5,4.0){\makebox(0,0)[r]{$S_2$}}
\end{picture}

\caption{The sequences of the two DNA molecules shown encode the same
  protein, but due to the differences in expected codon geometry, one
  of them is noticeably more bent than the other.}
\label{fig:bendingdemo}
\end{figure}

\subsection{The Geometric Pressure Hypothesis}

Note that the last example of two DNA sequences with very different
expected geometries is in some ways similar to the opening example of
a simple duplexed code.  Just as Fred and Georgina can have different
interpretations of the same signal of numbers, a sequence of codons
can be interpreted either as encoding a protein or as influencing the
shape of the DNA molecule.

The shape of a DNA molecule is relevant to its biological function.
It must bend in the right places and be straight in the right places
in order for the enzymes and RNA responsible for transcription to be
able to occur.  This gives biological importance to the question of
how well-duplexed the genetic and geometric codes discussed in the
previous two sections are.  For instance, if they are very
poorly-duplexed, then it could often be the case that a DNA molecule
cannot encode the protein that a creature needs unless it bends in a
bad way.  Conversely, it would be to a creature's advantage for the
codes to be well-duplexed because then it would always be able to
simultaneously encode whatever protein and geometry are optimal.

The geometric code was not something that evolution could act
upon since it is determined by the laws of physics and chemistry.
However, as we have seen, the genetic code could have been different
and likely was influenced by natural selection.  In \cite{Kasman}, it
was hypothesized that one of the factors that influenced that
selection was pressure to ensure that the genetic and geometric codes
were well-duplexed.

\section{How Well Duplexed is the Real Genetic Code as Compares with Alternatives?}

To test the ``Geometric Pressure Hypothesis'' (GPH), two different
measures of duplexing efficiency were developed in \cite{Kasman}.
Then, the duplexing efficiency of the geometric code with the natural
genetic code was compared with its average duplexing efficiency with a
large set of reasonable alternative genetic codes.

\subsection{Alternative Genetic Codes}

Let $\sym$ be the group of permutations on the set $\codonset$ of codons.  Then for any $\sigma\in \sym$, 
$\Phi_{\sigma}=\Phi_I\circ \sigma$ is the function from $\codonset$ to
$\aminoset$ which first replaces the codon $c\in\codonset$ with its
image $\sigma(c)$ and then applies the natural genetic code function
$\Phi_I$ to that.  (In other words, the function $\Phi_{\sigma}$ would be represented by a table very much like Table~\ref{tbl:geneticcode} for the natural genetic code above, but the codons would be rearranged according to the permutation $\sigma$.)

Notice that no matter which permutation $\sigma$ is selected, the
alternative genetic code $\Phi_{\sigma}$ has in common with the
natural genetic code $\Phi_I$ not only that it is a map from
$\codonset$ to $\aminoset$ but also that for each amino acid
$a\in\aminoset$ the preimages are of the same size:
$$
\ |\Phi_I^{-1}(a)|=|\Phi_{\sigma}^{-1}(a)|.
$$

However,  not all of those alternative genetic codes are realistic.  For most choices of permutation $\sigma$, the alternative
genetic code $\Phi_{\sigma}$ will not have the property that two
codons are more likely to encode the same amino acid (or chemically
similar amino acids) when their first two bases are equal, which will have already noted is a property of the real genetic code which has evolutionary advantages.  Since we only want to consider alternative codes that also have this property, the
permutations considered in \cite{Kasman} were further restricted: we
considered not arbitrary permutations but only ones with the property
that the first two bases in the codons $\sigma(c_1)$ and $\sigma(c_2)$
are equal if and only if the first two bases of $c_1$ and $c_2$ are
equal.  Let $S$ be the set of such permutations.
Symbolically, we can define the restricted set $S\subset \sym$ of permutations using the map $d(b_1b_2b_3)=b_1b_2$ which projects a codon onto its initial dimer as follows:
$$
S=\left\{\sigma\in \sym\ :\  \forall c,c'\in\codonset\ d(\sigma(c))=d(\sigma(c'))\Leftrightarrow d(c)=d(c')\right\}.
$$

To test the GPH in \cite{Kasman}, the duplexing efficiency of the natural genetic code
with the geometric code was compared with the expected value for the duplexing efficiency
of the alternatives indexed by the set $S$.
Since there are 
$(4^2)!(4!)^{16}$ permutations in $S$, this is a
very large set of permutations to consider.

\subsection{Total Network Length}

The natural genetic code is shown in Table~\ref{tbl:geneticcode} and
the expected geometries of the codons is shown in 
Figure~\ref{fig:codonspread}.  
One way to combine this information is to draw an edge on the figure
between any two codons that encode the same amino acid, turning it
into a graph with vertices and edges.  Thus, for instance, an edge
would be drawn between the vertices labeled \basefont{TAT} and
\basefont{TAC} because they both encode the amino acid \basefont{Y},
while the vertex labeled \basefont{ATG} would not be connected to any
other vertices.

Each connected component of the graph corresponds to an amino acid.
If there are only short edges or no edges in the component associated
to an amino acid that you wish to encode in a sequence, then that
means you have almost no geometric choice in the DNA molecule's
expected geometry at that point.  On the other hand, if there are long
edges in the connected component, then you would have a choice of
different codons that encode that same amino acid but which would
cause a very different geometric configuration of the DNA molecule.  

One could do the same for an alternative genetic code $\Phi_{\sigma}$.  That graph would have the same number of edges as the graph for the natural genetic code, but they would not have the same lengths.  

With all of this in mind, we define the ``total network length'' of the  genetic
code $\Phi_{\sigma}$ for any $\sigma\in S$ to be the sum of the lengths of the edges in the
graph\footnote{The second sum is over all distinct, unordered pairs in the pre-image $\Phi_{\sigma}^{-1}(a)$ and the length denotes the ordinary metric on $\R^3$ (i.e. $|(a,b,c)|=\sqrt{a^2+b^2+c^2}$).}:
$$
T_{\sigma}=\sum_{a\in\aminoset}\left(\sum_{c,c'\in\Phi_{\sigma}^{-1}(a)}\left|\bar{g}(c)-\bar{g}(c')\right|\right).
$$

In the case of the natural genetic code, $T_I$ was found to be $45.4238$ \AA.
 Because a larger total network length indicates more geometric freedom within the genetic code, if the
GPH was true, one might expect the total network length for the
natural genetic code to be  \textit{large} as compared with the total
network length for the alternative codes which are not found in nature.

Disappointingly, that is not what was found in \cite{Kasman}.  A 95\% confidence
interval was constructed for the expected value of the total network length 
over all permutations in $S$.  It was found that the average
total network length $E_{S}(T_{\sigma})$ is probably between $45.7655$ and $45.9639$
\AA.  If so, then 
$$
45.4238=T_I<45.7655<E_{S}(T_{\sigma})<45.9639
$$
Contrary to the predictions of the GPH, the
total network length $T_I$ for the natural genetic code is apparently a bit
\textit{smaller} than average rather than being especially large.

\subsection{Mutual Information with a Discretized Geometric Code}

The previous paper \cite{Kasman} also uses the concept of mutual information to quantify the mutual information of DNA's geometric and genetic codes.  Using mutual information as a measure of duplexing efficiency has two big advantages over the use of total network length as described in the previous section:
\begin{itemize}
\item Firstly, it is a well-known measure of duplexing efficiency which is widely studied and used, whereas total network length is an \textit{ad hoc} approach developed only for this particular project.
\item Total network length was based only on the expected values in Table~\ref{tbl:olsondata} and therefore ignored the standard deviations that represented the flexibility of the dimers.  Of course, once that flexibility is taken into account, the ``geometric code'' is no longer a function since there is more than one possible configuration for each codon.
Because the definition of mutual information in \ref{eqn:MI} involves probabilities, it is well suited to address this situation.
\end{itemize}

In \cite{Kasman}, the geometry of a codon was represented by a point in $\R^6$ where the first three numbers (like the image of $\bar{g}$ above) indicate the location of the center of the last ``rung'' of the codon relative to the first and the other three were angles indicating how it was tilted and twisted relative to the first.  Then, $\R^6$ was divided into $4096(=4^6)$ subsets called  `bins'.  If each bin is indexed by an element of $\Y=\{1,2,.\ldots,4095,4096\}$ then the geometric information is encoded into a map $g:\codonset\to\Y$.

Unlike any of the maps discussed earlier, $g$ is not a \textit{function} since given codon can be in many different possible geometric configurations due to its flexibility.  Although it is more likely to be in certain configurations than others, and so $g(c)$ is a random variable for any given codon $c$.  In order to compute the mutual information of the genetic codes $\Phi_{\sigma}$ with this geometric map $g$, we need to be able to compute the associated probabilities.  In \cite{Kasman} that was done by running a computer program which looped through a large number of different configurations and recorded the number of the `bin' in which they ended up.  In other words, the probabilities were computed \textit{empirically}, using the assumption that the dimer step parameters for are \textit{normally} distributed with the mean and standard deviation shown in Table~\ref{tbl:olsondata}.

Using this information it is now possible to compute (or, perhaps it would be better to say ``approximate'') the mutual information of any of the genetic codes $\Phi_{\sigma}$ with this geometric code $g$.  When this was done in \cite{Kasman}, was found that the mutual information of the \textit{natural} genetic code $\Phi_I$ with the geometric code $g$ is about $M(\Phi_I,g)\approx0.154144$ bits.  However, when the same computation was repeated for randomly selected alternative genetic codes $\Phi_{\sigma} $ for $\sigma\in S$ a 95\% confidence interval found that the average mutual information is probably between $0.14286$ and $0.148569$.  It it is, then
$$
0.14286\leq E_{S}(M(\Phi_{\sigma},g))\leq 0.148569<M(\Phi_I,g)\approx0.154144.
$$
Since a \textit{smaller} mutual information (closer to the ideal value of $0$) represents better duplexed codes, this means that the duplexing of the real genetic code $\Phi_I$ is \textit{worse} than average.  This, again, is the opposite of what would have been predicted by the GPH.

\section{New Results: Mutual Information via a Monte-Carlo style discretization}\label{sec:new}

Since the geometric parameters take values in a continuous space, the  functions $P_{f,g}$ and $P_g$ which appear in the formula for the mutual information are actually probability distribution functions whose values only become probabilities when integrated over regions of that space.  In the previous paper this computation was discretized in a \textit{rigid} way by ``binning'' the data into fixed and pre-determined subsets of equal size.

That approach
is plausible, but not uniquely so.
A more standard approach in numerical analysis is to consider a discretization based on choosing a suitable \textit{random} sample $Y_{N}$ of $N$ geometries that are chosen taking into account the Gaussian distributions in Table~\relax{\ref{tbl:olsondata}},  
and replace the usual discrete mutual information by
\[
M_{N}(f, g) = \frac{1}{W} \sum_{y \in Y_{N}} \sum_{a} P_{f, g}(a, y) \log_2 \left( \frac{P_{f, g}(a, y)}{P_f(a) P_g(y)} \right)\hbox{ with }
W = \sum_{y \in Y_{N}} \sum_{a} P_{f, g}(a, y),
\]
where $y$ is a codon geometry, $a$ is an amino acid, $f=f_{\sigma}$ is one of the genetic codes mapping codons to amino acids.
The normalization by factor $1/W$ is 
the volume element for approximate integration over the probability density $\sum_{a} P_{f, g}(a, y)$.
It normalizes the sum appropriately, in the sense of $M_N(f,g)$ approaching a common value $M(f,g)$ as $N$ increases.

The randomness of the sample $Y_N$ is one of the main differences between the prior results and this new approach.  Another difference is the randomness used in the approximation of the values of the probability distributions themselves.  Unlike the previous approach in which the probabilities were estimated by using rigidly chosen deviations from the expected values, this time a Monte Carlo approach will be utilized.  In particular, here we construct random choices for $N = 64n$ geometry samples as the union of a set of $n$ sample points for each codon, with the samples for each codon constructed from $n$ sets of random values for the twelve Hassan-Calladine parameters for the two dimers; 
the randomness based on assuming that these parameters are all independent and that each is normally distributed with mean and standard deviation as in Table~\relax{\ref{tbl:olsondata}}.  

The final, and perhaps most interesting, difference between the previous approach and the one taken in this section is an inversion of the geometric data which directly computes the probability that a given codon will take on a given geometric conformation.
The basic quantity needed is the probability distribution $P_d(\sigma)$ for the values of the dimer step $\sigma$ for dimer $d$ with
Hassan-Calladine parameters $\Delta_1, \Delta_2, \Delta_3, \theta_1, \theta_2, \theta_3$.
As above, this is based on the assumption that these parameters are independent and each is normally distributed, so:
\begin{equation} \label{P_d}
P_d(\sigma)=\prod_{i=1}^3
\frac{1}{\hat\Delta_i(d)\sqrt{2\pi}}\textup{exp}\left(
-\frac{(\Delta_i-\bar\Delta_i(d))^2}{2\hat \Delta_i(d)^2}
\right)
\prod_{j=1}^3
\frac{1}{\theta_j(d)\sqrt{2\pi}}\textup{exp}\left(
-\frac{(\theta_j-\bar\theta_j(d))^2}{2\hat \theta_j(d)^2}
\right)
.
\end{equation}
Consider codon $C$ consisting of dimers $d_1$ and $d_2$,
and for a choice of their dimer steps $\sigma_1$ and $\sigma_2$, denote the resulting codon geometry $y$ as the ``step product'' $\sigma_1 * \sigma_2$.
Then $P_C(y)$, the probability density of the codon $C$ having geometry $y$, is given by an integral over all paths to $y$:
\[
P_C(y) = \int_{\sigma_1} \left[ \sum_{\sigma_2 | \sigma_1 * \sigma_2 = y} P_{d_2}(\sigma_2) \right] P_{d_1}(\sigma_1) \,d\sigma_1
\]
For each pair of values for $y$ and $\sigma_1$, we must solve for all possible values $\sigma_2$;
fortunately this can be done explicitly, and generically there are only two such paths; this is detailed in the next subsection.

For each path the quantity $P_{d_2}(\sigma_2)$ will be evaluated using Eq.~\eqref{P_d}.
The outer integral over a six dimensional space is instead dealt with by a Monte-Carlo method:
noting that it is the integral w.r.t. a probability measure, we approximate by choosing a sample of $m$ random sets of values for the Hassan-Calladine parameters, in turn determining a set $\Sigma_{m}$ of values for $\sigma_1$, and averaging:
\[
P_C(y) \approx \frac{1}{m} \sum_{\sigma_1 \in \Sigma_m} \sum_{\sigma_2 | \sigma_1 * \sigma_2 = y} P_{d_2}(\sigma_2) 
\]

Then we assemble the pieces:
\[
\begin{split}
P_g(y) &= \frac{1}{64} \sum_{C } P_C(y)
\\
P_{f, g}(y) &= \frac{1}{64}\sum_{C | f(C) = a} P_C(y)
\\
& \text{and the easy one, the fraction of codons that give a specified amino acid:}
\\
P_f(a) &=  \frac{1}{64} \big | \{ C| f(C) = a \} \big |
\end{split}
\]

\subsection{Reconstructing the second dimer step}\label{second step}

What remains is to solve $\sigma_1 * \sigma_2 = y$ for the second dimer step $\sigma_2$;
that is, find the corresponding Hassan-Calladine angles $\theta_i$ and then the lengths $\Delta_i$.

Except for one case noted 
below (of negligible probability), the angles $\theta_i$, $i=1$--$3$ are determined up to negation of the pair $\theta_1, \theta_2$.
This comes from the formula
\begin{equation}
T = M R_3(\theta_3/2-\phi) R_2(\eta) R_3(\theta_3/2+\phi)
\end{equation}
as in Section 2.2 of \cite{Kasman}
(See also Eq's (9) of \cite{HC}.) 
The matrices
$M$ and $T$
are the frames respectively for the end of the first dimer and the end of the codon,
$R_2$ and $R_3$ are the familiar matrices for rotations about $y$ and $z$ axes
\[
R_2(\eta) = \begin{bmatrix} \cos\eta & 0 & -\sin\eta \\ 0&1&0 \\ -\sin\eta & 0 & \cos\eta \end{bmatrix}
\quad
R_3(\theta) = \begin{bmatrix} \cos\theta & -\sin\theta & 0 \\ \sin\theta & \cos\theta & 0 \\ 0&0&1 \end{bmatrix},
\]
and
$\eta = \sgn(\theta_2) \sqrt{\theta_1^2+\theta_2^2}$,
$\sin\phi=\theta_1/\eta$ with $-\pi \leq \phi \leq \pi$,

\medskip

From this,
\[
\theta_2 =  \sgn(\theta_2) \sqrt{\eta^2-\theta_1^2} = \eta \cos\phi
\]

Let $R=M^{-1}T = R_3(\theta_3/2-\phi) R_2(\eta) R_3(\theta_3/2+\phi)$ be the combined rotation.
\[
R_{33} = (R_3(\theta_3/2-\phi)^T e_3)^T R_2(\eta) (R_3(\theta_3/2-\phi) e_3)
= e_3^T  R_2(\eta) e_3
= (R_2(\eta))_{33}
= \cos\eta,
\]
so
\[
\eta = \pm \arccos R_{33} \in [0, \pi]
\]

\textbf{Case 1 (Generic): $-1<R_{33}<1$, so $0 < |\eta| < \pi$.}
\\
Defining $\alpha=\theta_3/2+\phi$ and $\beta = \theta_3/2-\phi$,
$\alpha$ is determined in $(-\pi,\pi]$ by
\\
\[
\cos\alpha = -R_{31}/\sin\eta, \quad\sin\alpha = R_{32}/\sin\eta
\]
and likewise $\beta$ by
\[
\cos\beta = R_{13}/\sin\eta, \quad\sin\beta =  R_{23}/\sin\eta
\]
$\sin\eta \neq 0$, so no problems here.

Then $\theta_3 = \alpha+\beta$,
$\phi = (\alpha-\beta)/2$,
$\theta_1=\eta \sin\phi$,
$\theta_2 = \eta \cos\phi $.

The two choices for $\eta$ likewise negate $\theta_1$ and $\theta_2$, but only shift $\theta_3$ by an irrelevant increment of $2\pi$.

%

\textbf{Case 2. $R_{33}=1$, so $\eta = 0$.}
\\
$\theta_1=\theta_2=0$, and $R = R_3(\theta_3)$ so $\theta_3$ is determined easily.

\textbf{Case 3: $R_{33}=-1$, so $\eta = \pm\pi$.}
\\
This is the problem case, as $R$ now depends only on $\phi$, not $\theta_3$, so the latter is not constrained at all.
However, the value $\eta=\pm\pi$ is extremely unlikely:
$\eta^2 = \theta_1^2+\theta_2^2$, and as seen in Table~\relax{\ref{tbl:olsondata}}, the values for the later two angles are far too small.

\medskip

Reconstructing the remaining Hassan-Calladine parameters $\Delta_i$ is now straightforward; they are related to the known positions of the ends of each dimer and the  Hassan-Calladine by a system of linear equations, as seen in the formulas
\[
T = (\vec v_1^{\top}\  \vec v_2^{\top}\ \vec v_3^{\top})=MR_3\left(\frac{{\theta_3}}{2}-\phi\right)R_2\left(\frac{\eta}{2}\right)R_3\left(\phi\right).
\]
with the (known) positions $\vec p_2$ and $\vec p_3$ of the ends of the second and third bases related by
\[
\vec p_3 = \vec p_2 + \Delta_1 \vec v_1 + \Delta_2 \vec v_2 + \Delta_3 \vec v_3
\]
from \cite{Kasman}; see also Eq's (10,11) of \cite{HC}.

\subsection{Numerical Results}\label{method B results}

The most accurate calculation so far for the true genetic code is with $N=4096$ samples and $m=2048$ samples for each evaluation of $P_C(y)$.
This gives $M_N(f_I, g) \approx 0.1834$, with standard error of the mean estimated at $0.003$.

Comparisons to  alternative genetic codes have been done with 16 randomly generated codes, each with $N=2048$ and $m=1024$;
the random codes then give a mean $E_S(M_N(f_{\sigma}, g))$ value of $0.1776$ with
95\% confidence interval
$[ 0.1708, 0.1844]$.
With those same sample size parameters for the true genetic code the result was $0.1877$.
Much as seen in Section 3.3, this is slightly out of the 95\% confidence interval,
in the opposite direction to that suggested by the Geometric Pressure Hypothesis:
$$
0.1708\leq E_S(M_N(f_{\sigma}, g))\leq  0.1844< M_N(f_I,g)\approx 0.1877.
$$

\section{Conclusions}

A given genetic sequence can be interpreted as encoding a protein and also as influencing the geometry of the DNA molecule that carries it.  This is therefore a situation like the one in the Introduction where Georgina and Fred are each interpreting the same signal differently.  It is therefore of interest to understand how well-duplexed these two ``codes'' are.

Combining the results of the previous paper \cite{Kasman} with the new results in Section~\ref{sec:new} this has now been done in three different ways.  Disappointingly, each time the answer has turned out to be ``about what you'd expect if the real genetic code was just selected randomly, but maybe a little worse''.  In other words, contrary to the predictions of the Geometric Pressure Hypothesis (GPH), the natural genetic code does not appear to be especially well-duplexed.  In other words, even if one replaced the natural genetic code with a random alternative, it is likely that there there would be more freedom in the geometry of the DNA molecule while encoding any given protein.

 It is interesting to speculate about what that tells us about the evolution of the genetic code.  On the one hand, it could simply imply that there is not much evolutionary advantage in having the ability to alter the shape of the DNA molecule without changing the protein it encodes.  However, that is not the only possible explanation.  Another intriguing possibility, which was raised during the discussion after the talk in the session for which this volume serves as the proceedings, is that the code became fixed before the molecules became large enough for it to matter.  In particular, it does seem likely that the geometry of the molecule is not very important when the chromosome is very short.  So, if the genetic code that we are familiar with was shaped during an early period in evolution when the genome of living creatures were all very small, the GPH might not have applied.  And, since the genetic code is no longer very malleable (as demonstrated by its near ubiquity), it might no longer have been able to change once the molecules grew large enough for their geometry and topology to matter.
 
In any case, whatever the explanation may be, the new computations have only re-confirmed the answer found previously to the question of the title.  Since the natural genetic code appears only slightly less well-duplexed with the geometric code than an average alternative, it does not appear that the evolution of the genetic code was shaped by any pressure to optimize it.

\bibliographystyle{amsplain}

\end{document}